\documentclass[12pt]{article}
\usepackage{scicite}
\usepackage{times}
\usepackage{graphicx}
\usepackage{bbm}
\usepackage{soul}
\usepackage{siunitx} 

\usepackage[draft]{todonotes}   
\usepackage{comment}

\newenvironment{sciabstract}{%
\begin{quote} \bf}
{\end{quote}}

\newcounter{lastnote}

\title{
Single-Photon Switching and Entanglement of Solid-State Qubits in an Integrated Nanophotonic System
}

\author{
A. Sipahigil$^{1,\ast}$, R. E. Evans$^{1,\ast}$, D. D. Sukachev$^{1,2,3,\ast}$, M. J.  Burek$^{4}$, J. Borregaard$^1$, \\ M. K. Bhaskar$^1$, C. T. Nguyen$^1$, J. L. Pacheco$^{5}$, H. A. Atikian$^{4}$, C. Meuwly$^{4}$,\\ R. M. Camacho$^{5}$, F. Jelezko$^{6}$, E. Bielejec$^{5}$, H. Park$^{1,7}$, M. Lon\v{c}ar$^4$, M. D. Lukin$^{1,\dagger}$\\
\\
\baselineskip12pt 
\normalsize{$^{1}$Department of Physics, Harvard University, Cambridge, MA 02138, USA}\\
\normalsize{$^{2}$Russian Quantum Center, Skolkovo, Moscow 143025, Russia}\\
\normalsize{$^{3}$P.\,N. Lebedev Physical Institute of the RAS, Moscow 119991, Russia}\\
\normalsize{$^{4}$John A. Paulson School of Engineering and Applied Sciences,}\\
\normalsize{Harvard University, Cambridge, MA 02138, USA}\\
\normalsize{$^{5}$Sandia National Laboratories, Albuquerque, NM 87185, USA}\\
\normalsize{$^{6}$Institute for Quantum Optics, University of Ulm, 89081 Ulm, Germany}\\
\normalsize{$^{7}$Department of Chemistry and Chemical Biology,}\\
\normalsize{Harvard University, Cambridge, MA 02138, USA}\\ 
\normalsize{$^\ast$ These authors contributed equally to this work.}\\
\normalsize{$^\dagger$ Corresponding author. E-mail:  lukin@physics.harvard.edu}
}

\date{}

\begin{document} 
\baselineskip16pt 
\maketitle 

\begin{sciabstract}
Efficient interfaces between photons and quantum emitters form the basis for quantum networks and enable nonlinear optical devices operating at the single-photon level. 
We demonstrate an integrated platform for scalable quantum nanophotonics based on silicon-vacancy (SiV) color centers coupled to nanoscale diamond devices.
By placing SiV centers inside diamond photonic crystal cavities, we realize a quantum-optical switch controlled by a single color center.  
We control the switch using SiV metastable orbital states and verify  optical switching at the single-photon level by using photon correlation measurements. 
We use Raman transitions to realize a single-photon source with a tunable frequency and bandwidth in a diamond waveguide. Finally, we create entanglement between two SiV centers by detecting indistinguishable Raman photons emitted into a single waveguide. Entanglement is verified using a novel superradiant feature observed in photon correlation measurements, paving  
the way for the realization of quantum networks.
\end{sciabstract}

\noindent Efficient interfaces between photons and quantum emitters are central to applications in quantum science \cite{kimble2008quantum, chang2014quantum,ladd2010quantum} but are challenging to implement due to weak interactions between single photons and individual quantum emitters. 
Despite advances in the control of microwave and optical fields using cavity and waveguide quantum electrodynamics (QED) to achieve strong interactions\cite{tiecke2014nanophotonic,reiserer2014quantum,englund2007controlling,Javadi2015,sun2016quantum,mlynek2014observation}, the realization of integrated quantum devices where multiple qubits are coupled by optical photons remains an outstanding challenge\cite{lodahl2015interfacing}. 
In particular, due to their complex environments, solid-state emitters have optical transitions that generally exhibit a large inhomogeneous distribution\cite{badolato2005deterministic, lodahl2015interfacing}, rapid decoherence\cite{sun2016quantum} and significant spectral diffusion, especially  in nanostructures\cite{faraon2012coupling}. 
Moreover, most solid-state emitters appear at random positions, making the realization of scalable devices with multiple emitters difficult \cite{badolato2005deterministic, riedrich2014deterministic}.

\noindent \textbf {Diamond platform for quantum nanophotonics. }
Our approach uses negatively-charged silicon-vacancy (SiV) color centers integrated into diamond nanophotonic devices.  
The SiV center is a point defect in diamond\cite{hepp2014electronic} with remarkable properties: its optical transitions are consistently nearly lifetime-broadened, and the inhomogeneous (ensemble) distribution of SiV transition frequencies can be on the order of the lifetime-broadened linewidth in high-quality bulk diamond\cite{rogers2014multiple}. 
These properties arise from the inversion symmetry of the SiV center which results in a vanishing permanent electric dipole moment for the SiV orbitals, dramatically reducing their sensitivity to electric fields\cite{sipahigil2014indistinguishable}.
The optical transitions are therefore protected from charge fluctuations that cause large inhomogeneous distributions and spectral diffusion\cite{faraon2012coupling,Javadi2015}. 
These properties allow SiV centers to be integrated into nanostructures while maintaining their coherence properties\cite{evans2016narrow}. 

\begin{figure}
\begin{center}
		\includegraphics[width=\linewidth]{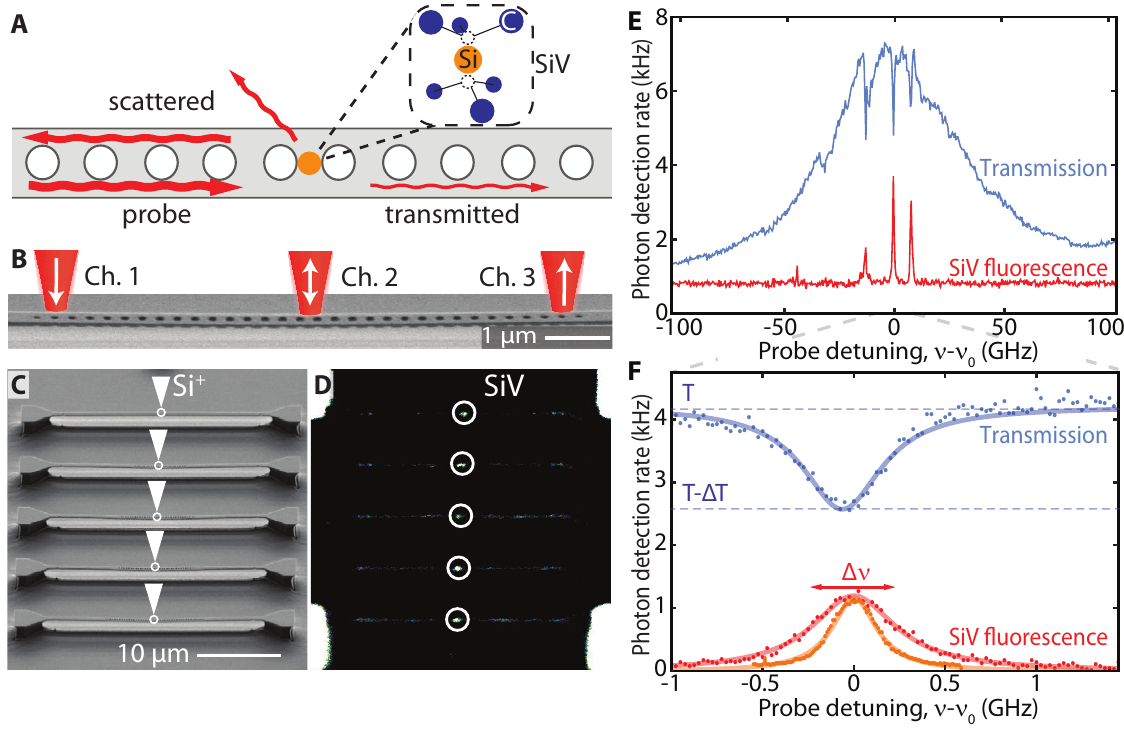}
\end{center}
		\caption{
\textbf{Positioning and strong coupling of SiV centers in diamond cavities.} 
\textbf{(A)} Schematic of a single SiV center integrated into a diamond nanophotonic crystal cavity. Inset: SiV molecular structure. 
\textbf{(B)} Scanning electron micrograph (SEM) of a nanophotonic crystal cavity. 
Three optical beams are used to excite the waveguide mode (Ch.~1), to detect fluorescence scattering and to control the SiV (Ch.~2) and to detect transmission (Ch.~3).  
\textbf{(C)} SEM of five cavities fabricated out of undoped diamond. After fabrication, SiV centers are deterministically positioned at the center of each cavity using focused Si${}^+$ ion beam implantation. 
\textbf{(D)} SiV fluorescence is detected at the center of each nanocavity shown in \textbf{(C)}, indicating high SiV-nanocavity device creation yield. 
\textbf{(E)} Cavity transmission (blue, Ch.~3) and SiV scattered fluorescence (red, Ch.~2) are recorded as the probe laser frequency $\nu$ is scanned across the SiV resonance at $\nu_0=406.706 \,$\si{\THz}. Three peaks correspond to three SiV centers resonantly coupled to the cavity, resulting in suppressed transmission at the corresponding frequencies. 
\textbf{(F)} Strong extinction, $\Delta T/T=38(3)\%$, of probe transmission from a single SiV center. Optical transition linewidths $\Delta \nu$ are measured with the cavity detuned (orange) and on resonance (red, count rate offset by \SI{-150}{Hz} and multiplied by 3.3 for clarity) with the SiV transition. On resonance, the transition is radiatively broadened from $298(5)$ to $590(30)$\,\si{\MHz}. \textbf{(F)} is a zoomed-in version of \textbf{(E)} around $\nu-\nu_0=0$.
}
	\label{fig:cavity}
\end{figure}

The stable quantum emitters are integrated into one-dimensional diamond waveguides and photonic-crystal cavities with small mode volumes ($V$) and large quality factors ($Q$).
These nanophotonic devices are fabricated using angled reactive ion etching to scalably create free-standing single-mode structures starting from bulk diamond\cite{burek2014high,burek2012free,SOM}.  As an example, Fig.~1C shows structures 
consisting of an anchor for mechanical support, a notch for free space-waveguide coupling, a waveguide section on each side and a cavity (Fig.~1B) defined by a tapered set of holes. 
The measured cavity $Q=7200\,(500)$ is limited predominantly by
photon decay  to the waveguide, so the system has high transmission on resonance.
We measure the cavity mode profile and infer $V\sim 2.5(\lambda /n)^3$ using a high and uniform density SiV ensemble (Fig.~S4).%
To obtain  optimal coupling between an individual SiV center and the cavity mode, we use a focused ion beam to implant Si${}^+$ ions at the center of the cavities as illustrated in Fig.~1C. 
To form SiV centers and mitigate crystal damage from fabrication and implantation, we subsequently anneal the sample at \SI{1200}{\celsius} in vacuum\cite{evans2016narrow}.
This targeted implantation technique enables positioning of the emitters inside the cavity with close to \SI{40}{\nano\meter} precision in all three dimensions\cite{SOM} and control over the isotope and average number of implanted Si${}^+$ ions. 
Unlike approaches where nanophotonic structures are fabricated around pre-characterized or randomly-positioned solid-state emitters\cite{badolato2005deterministic, riedrich2014deterministic}, we fabricate around $2000$ SiV-cavity nodes on a single diamond sample with optimal spatial alignment. 
For example, Fig.~1D shows the fluorescence image of the array of cavities implanted with Si${}^+$ ions in Fig.~1C. SiV fluorescence is detected at the center of each cavity, demonstrating high-yield creation of SiV-cavity nodes. 
The number of SiV centers created at each spot varies due to the Poisson distribution of the number of implanted Si${}^+$ ions and the approximately $2\%$ conversion yield from Si${}^+$ to SiV\cite{SOM}.
For our experiments, we create an average of $\sim 5$ SiV centers at each cavity.
Because the individual SiVs can be subsequently resolved in the frequency domain, this results in nearly deterministic device creation such that the majority of SiV-cavity nodes on our chip can be used for the experiments described below.
We characterize the coupled SiV-cavity system at \SI{4}{\kelvin} in a helium cryostat\cite{SOM}. 
As shown in Figs.~1B and S1, three optical beams are focused on the nanostructure 
to excite the waveguide mode (Channel~1), to detect fluorescence scattering and to control the SiV (Channel~2) and to detect transmission (Channel~3). In subsequent experiments (Figs.~4 and 5), efficient collection through tapered optical fiber is employed.
We scan the frequency $\nu$ of the weak excitation laser 
across the SiV resonance $\nu_0$ and monitor the transmitted and scattered field intensities (Fig.~1E).
We observe three fluorescence peaks in Channel~2 from three SiV centers in a single cavity (red curve in Fig.~1E)\cite{SOM}. 
At the same time,  
within the broad 
cavity transmission spectrum measured in Channel~3, each of these three resonances results in strong extinction of the cavity transmission indicating that all three SiV centers couple to the cavity mode. 
The strength of the SiV-cavity coupling can be directly evaluated using the data presented in Fig.~1F. When the cavity is off-resonant with the emitter, the transition linewidth is $\Delta\nu=298\,(5)$\,\si{MHz} (Fig.~1F, yellow curve) and the excited state has a lifetime of $\tau_e = 1.8\,(1)$\,\si{ns}. 
This is close to the lifetime-broadening limit of $\SI{90}{MHz}$ with additional nonradiative broadening likely due to a combination of finite-temperature effects\cite{jahnke2015electron} and residual spectral diffusion\cite{evans2016narrow}. 
When the cavity is tuned into resonance\cite{SOM}, the optical transition linewidth is radiatively broadened to $\Delta\nu=590\,(30)$\,\si{MHz} (Fig.~1F, red curve) with a corresponding measured reduction in lifetime $\tau_e = 0.6\,(1)$\,\si{ns} (limited by detection bandwidth).
At the same time, we find that a single SiV results in $38\,(3)\%$ extinction of the probe field in transmission (Fig.~1F, blue curve). 
Based on the radiative broadening shown in Fig.~1F, we infer a cooperativity of $C=4g^2/\kappa \gamma= 1.0\,(1)$ for the SiV-cavity system with cavity QED parameters $\{g, \kappa, \gamma\}/2\pi=\{2.1,57,0.30\}$\,\si{GHz} where $g$ is the single-photon Rabi frequency, $\kappa$ is the cavity intensity decay rate, and $\gamma$ is the SiV optical transition linewidth\cite{SOM}. 
This cooperativity estimate is consistent with the measured transmission extinction when we account for effects associated with the multi-level structure of the SiV (see below).
\noindent \textbf{Quantum-optical switch based on a single SiV center.} The coupled emitter-cavity system can be used to create strong interactions between single photons\cite{chang2007single, chang2014quantum}. 
When combined with additional metastable states, it can enable single-photon switching and quantum logic between single photons and SiV centers\cite{duan2004scalable,imamoglu1999quantum}. 
To probe the nonlinear response of the SiV-cavity system, we repeat the transmission and linewidth measurements of Fig.~1F at increasing probe intensities.  As expected \cite{rice1988single,chang2007single}, we find that the system response saturates at a level less than a single photon per Purcell-enhanced excited-state lifetime (Fig.~3A), resulting in power broadening in fluorescence ($\Delta\nu$) and reduced extinction in transmission ($\Delta T/T$) \cite{SOM}. 
%

\begin{figure}
\begin{center}
		\includegraphics[width=\linewidth]{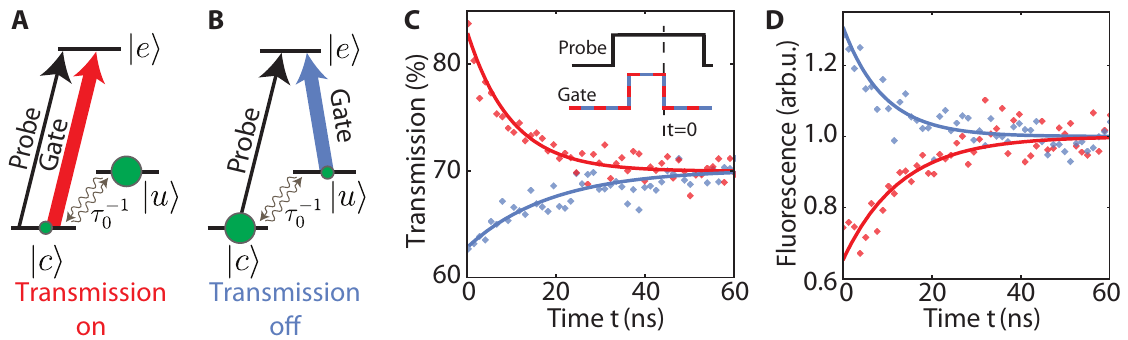}
\end{center}
		\caption{
		\textbf{An all-optical switch using a single SiV.} 
		The transmission of a probe field is modulated using a gate pulse that optically pumps the SiV to state $|u\rangle$ \textbf{(A)} or $|c\rangle$ \textbf{(B)}. The states $|u\rangle$ and $|c\rangle$ are metastable orbital states with the same spin projection, separated by \SI{64}{\giga\hertz} due to the spin-orbit interaction and crystal strain \cite{hepp2014electronic,jahnke2015electron}.
		\textbf{(C)} Probe field transmission measured after the initialization gate pulse. Initialization in state $|u\rangle$ ($|c\rangle$) results in increased (suppressed) transmission compared with the steady-state value.
		\textbf{(D)} The fluorescence response shows the opposite behavior, with pumping to state $|u\rangle$ ($|c\rangle$) resulting in suppressed (increased) fluorescence compared with the steady-state value.
		}
	\label{fig:switch}
\end{figure}

We realize an all-optical switch with memory by optically controlling the metastable orbital states\cite{pingault2014all,rogers2014all,becker2016ultrafast} of a single SiV (Fig.~2).
Specifically, we use a \SI{30}{\ns} long gate pulse to optically pump the SiV to an orbital state that is uncoupled ($|u\rangle$, Fig.~2A) or coupled ($|c\rangle$, Fig.~2B) to a weak probe field resonant with the cavity.
The $|u\rangle\leftrightarrow|e\rangle$ transition is detuned from the cavity and the probe field by \SI{64}{\GHz}.
The response of the system to the probe field after the gate pulse is monitored both in transmission (Fig.~2C) and in fluorescence (Fig.~2D). 
If the gate pulse initializes the system in state $|c\rangle$ (blue curves), the transmission is reduced while the fluorescence scattering is increased. 
Initializing the system in state $|u\rangle$ (red curves) results in increased transmission and reduced fluorescence scattering. 
The observed modulation demonstrates switching of a weak probe pulse by a classical gate pulse.
The switch memory time is limited by a thermal phonon relaxation process between $|c\rangle$ and $|u\rangle$ that depolarizes the system over $\tau_0\sim\ $\SI{10}{\ns} at \SI{4}{\kelvin}\cite{jahnke2015electron}.
In the steady state, we measure an approximately $65\%$ probability for the SiV to be in state $|c\rangle$, consistent with the expected thermal occupation.

To investigate the optical nonlinearity and switching behavior at the single-photon level, we resonantly excite the SiV-cavity system with a weak coherent state and measure photon statistics of the scattered and transmitted fields. 
To this end, scattered and transmitted light are each split to two detectors (Fig.~S1), 
allowing us to measure normalized intensity auto-correlations for the scattered ($g^{(2)}_{SS}(\tau)$, Fig.~3B) and transmitted ($g^{(2)}_{TT}(\tau)$, Fig.~3C) fields as well as cross-correlations between the two channels ($g^{(2)}_{ST}(\tau)$, Fig.~3D).
At short timescales determined by the excited state lifetime $\tau_e$, we observe strong antibunching of photons scattered by the SiV center  ($g^{(2)}_{SS}(0)=0.15(4)$), consistent with scattering from a single emitter.
In transmission, the photons are strongly bunched with $g^{(2)}_{TT}(0)=1.50(5)$. 
This photon bunching in transmission results from the interference between the weak probe field and the anti-bunched resonant scattering from the SiV. 
The destructive interference for single photons yields preferential transmission of photon pairs and is a direct indication of nonlinear response at the single-photon level\cite{rice1988single,chang2007single}. In other words, a single photon in an optical pulse switches a second photon, and the system acts as a photon number router where single photons are scattered while photon pairs are preferentially transmitted. 
Finally, both bunching ($g^{(2)}_{ST}(0)=1.16(5)$) and anti-bunching are observed for scattering-transmission cross-correlations at fast and slow timescales respectively\cite{SOM}. 
%
\begin{figure}
		\begin{center}
		\includegraphics[width=.9\linewidth]{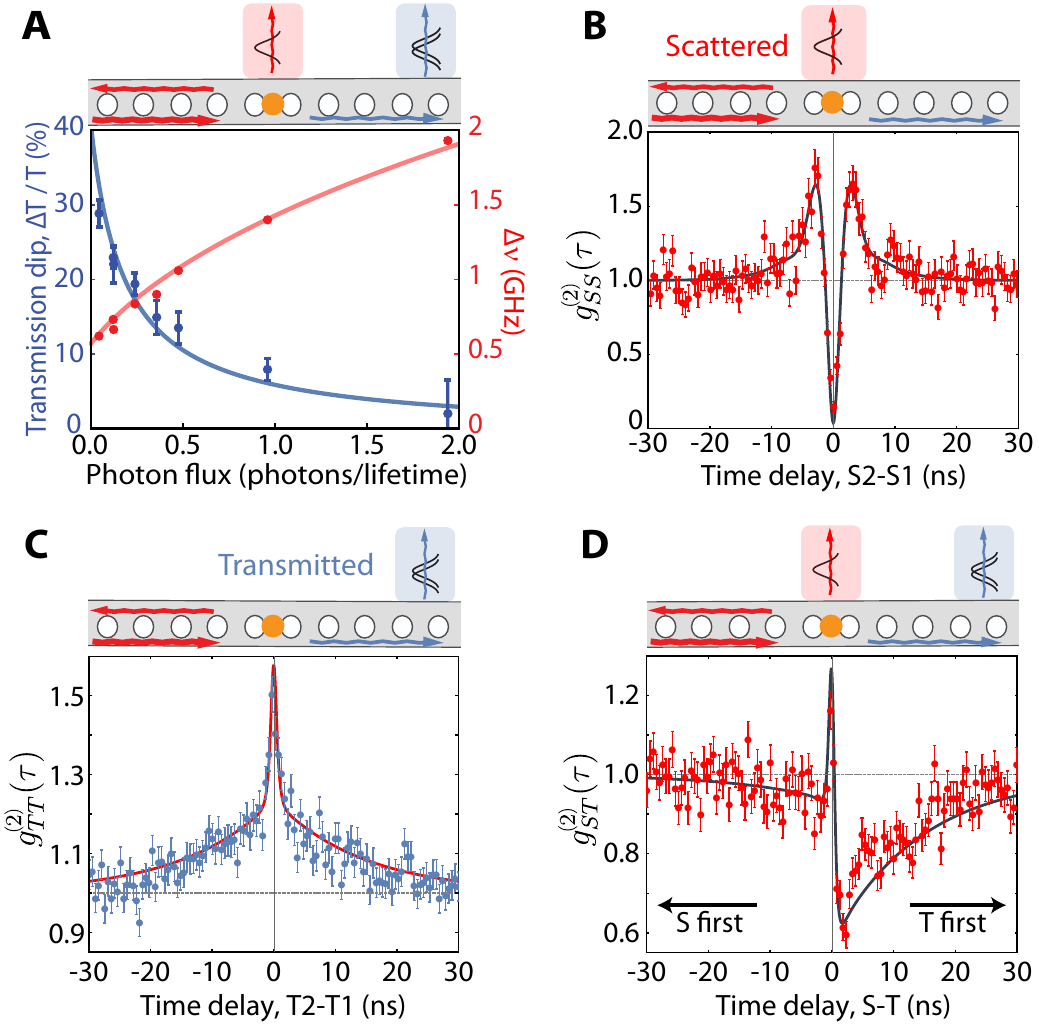}
		\end{center}
		\caption{
			\textbf{Single-photon switching.}
	\textbf{(A)} Cavity transmission and absorption linewidth as a function of probe photon flux. Atomic saturation at the single-photon level power-broadens fluorescence (increased $\Delta \nu$) and reduces the extinction $\Delta T/T$ in transmission\cite{SOM}. 
	\textbf{(B, C)} Normalized intensity autocorrelation function, $g^{(2)}(\tau)$, measured for the scattered (fluorescence) and transmitted fields. 
	The scattered field shows antibunching \textbf{(B)}, while the transmitted photons are bunched with an increased contribution from photon pairs \textbf{(C)}.
	\textbf{(D)} Intensity cross-correlation between the scattered and transmitted fields.
	The detection of a transmitted photon leaves the system in a different state than the detection of a scattered photon, resulting in the asymmetry of $g^{(2)}_{ST}(\tau)$.
	The system response at both the \SI{1}{\ns} and \SI{10}{\ns} timescales results from multilevel dynamics that are captured by our model (solid curves, see \cite{SOM}). The $g^{(2)}_{SS}$ is measured under different conditions with above saturation excitation\cite{SOM}. 
		}
	\label{fig:nonlinear}
\end{figure}

To understand the system saturation and switching responses in Figs.~2-3, we model the quantum dynamics of the SiV-cavity system using the cavity QED parameters measured in Fig.~1 and a three-level model of the SiV center
 as shown in Figs.~2 and S7. 
The results of our calculation \cite{SOM} are in excellent agreement with our observations (solid curves in Figs.~1-3). Specifically, the presence of a second metastable state, $|u\rangle$, reduces the extinction in linear transmission (Fig.~1F) and affects the nonlinear saturation response\cite{SOM,chang2007single}. 
The metastable state $|u\rangle$ also causes dynamics in the photon correlation measurements (Fig.~3) at the thermal relaxation timescale of $\tau_0 \sim 10\,\si{\ns}$ between $|u\rangle$ and $|c\rangle$.  
Upon detection of a transmitted  photon, the SiV has an increased likelihood to be projected into state $|u\rangle$, where the SiV is not excited by laser light. Consequently, for a time period ($\sim \tau_0$) given by the lifetime of state $|u\rangle$, the SiV-cavity system will have higher transmission and reduced scattering resulting in enhanced $g^{(2)}_{TT}$ and suppressed $g^{(2)}_{ST}$. 
If instead a scattered photon is detected first, the SiV is preferentially projected to the coupled state $|c\rangle$.
This state undergoes two-level dynamics associated with decreased scattering and enhanced transmission at short times on the order of the excited state lifetime $\tau_e$.
At longer times of order $\tau_0$ while the system is more likely to remain in the coupled state $|c\rangle$, it exhibits enhanced scattering ($g^{(2)}_{SS}$, Fig.~3A) and somewhat reduced transmission ($g^{(2)}_{ST}$, Fig.~3C).
These results are also consistent with the optical switching dynamics in Fig.~2 where the system is polarized using a classical gate pulse instead of a single photon detection event and the relaxation dynamics at the same timescale $\tau_0$ are observed.

\noindent \textbf{Tunable single-photon source using Raman transitions.} A key challenge for building scalable quantum networks using solid-state emitters is the spectral inhomogeneity of their optical transitions. Even though the inhomogeneous broadening of SiV centers is suppressed  by inversion symmetry, the SiV centers inside nanostructures still display substantial  inhomogeneous distribution (seen in Fig.~1E) due to the residual strain from device fabrication\cite{evans2016narrow}. 
To mitigate the effect of inhomogeneous broadening,  
we use Raman transitions  between the metastable orbital states of SiV centers. When  a single SiV center is excited from the state $|u\rangle$ by a  driving laser with  a detuning of $\Delta$, the emission spectrum includes a spontaneous component  at frequency $\nu_{ec}$ (labeled $S$ in Fig.~4B) and a Raman component at frequency $\nu_{ec}-\Delta$ (labeled $R$ in Fig.~4B). 
The Raman emission frequency and bandwidth can be manipulated by choosing, respectively,  the frequency and the intensity of the driving laser. 

\begin{figure}
\begin{center}
		\includegraphics[width=.8\linewidth]{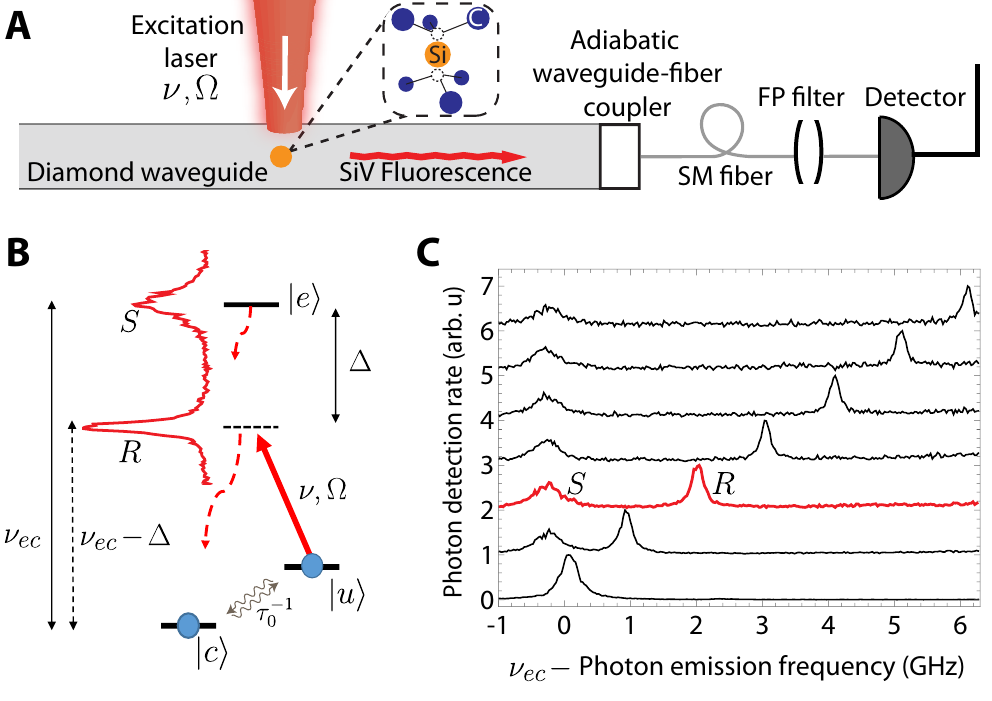}
\end{center}
		\caption{
	\textbf{Spectrally-tunable single-photons using Raman transitions.}
	\textbf{(A)} A single SiV center in a diamond waveguide is excited with a laser field from free space. Photons scattered into the diamond waveguide are coupled to a single-mode fiber using an adiabatic waveguide-fiber coupler with  $\geq70\%$ efficiency\cite{tiecke2015efficient,SOM}.  A scanning Fabry-Perot (FP) filter cavity ($\kappa_{FP}/2\pi=\SI{150}{MHz}$, $FSR=\SI{37}{GHz})$ is used as a high-resolution filter to measure emission spectra. 
\textbf{(B)} Under off-resonant excitation at a single photon detuning $\Delta$, the emission spectrum contains spontaneous emission (labeled $S$) at frequency $\nu_{ec}$ and narrow Raman emission (labeled $R$) at frequency $\nu_{ec}-\Delta$. The vertical frequency axis is not to scale.
	\textbf{(C)} 
	For the curves shown, the detuning $\Delta$ is varied from $0$ to \SI{6}{GHz} in steps of \SI{1}{GHz} and a corresponding tuning of the Raman emission frequency is observed. 
	The red curves in \textbf{(B)} and \textbf{(C)} are the same data.  
	}
	\label{fig:statistics}
\end{figure}

Tunable single-photon emission is realized experimentally by implanting SiV centers inside a one-dimensional diamond waveguide 
and exciting the emitters from free space using a continuous driving laser (Fig.~4A). The diamond waveguide has the same geometry as the cavity shown in Fig.~1B but lacks the periodically-patterned holes. The fluorescence scattering into the diamond waveguide is coupled to a tapered single-mode fiber with $\geq 70\%$ efficiency using adiabatic mode transfer\cite{tiecke2014nanophotonic,tiecke2015efficient,SOM}.
A scanning Fabry-Perot (FP) cavity is used as a high-resolution filter to measure fluorescence spectra and suppress scattered light from the driving laser and other transitions.
As we change the driving laser frequency from near-resonance up to a single-photon detuning of $\Delta=\SI{6}{GHz}$ in steps of \SI{1}{GHz}, we  observe a corresponding tuning of the Raman emission frequency $\nu_{ec}-\Delta$ while the spontaneous emission frequency remains nearly fixed at $\nu_{ec}$ up to an AC Stark shift (Figs.~4C and ~S8).

The observed linewidth of the Raman emission $R$ is below that of the spontaneous component $S$.
The Raman linewidth can be controlled by both the detuning and the power of the driving laser, 
and is ultimately limited by the ground state coherence between states $|u\rangle$ and $|c\rangle$ at large detunings. At large detunings and low power, we measure a subnatural Raman linewidth of less than \SI{30}{\MHz} \cite{SOM}. 
The nonclassical nature of the Raman emission is demonstrated via photon correlations measurements. We find that the Raman photons from a single SiV are antibunched with $g^{(2)}_{single}(0)=0.16\,(3)$ (orange curve in Fig.~5D) close to the ideal limit $g^{(2)}_{single}(0)=0$ \cite{SOM}. 
For the continuous excitation used here, we detect Raman photons at a rate of $\sim\SI{15}{kHz}$ from a single SiV center. 
After a Raman scattering event, the SiV cannot scatter a second photon within the metastable orbital state relaxation timescale $\tau_0$ which limits the Raman emission rate.
This rate can be improved using a pulsed excitation scheme in which the SiV center is first prepared in the state $|u\rangle$ via optical pumping and subsequently excited with a driving laser field of desired shape and duration. This approach can further be used to control the timing and the shape of the single-photon pulses\cite{kuhn2002deterministic}.  
Unlike previous demonstrations of Raman tuning of solid-state quantum emitters \cite{fernandez2009optically,he2013indistinguishable, sweeney2014cavity}, the tuning range demonstrated here\cite{SOM} is comparable to the inhomogeneous distribution of the SiV ensemble and can thus be used to
tune pairs of SiV centers into resonance.

\noindent \textbf{Entanglement of SiV centers in a diamond nanophotonic waveguide. }
 Quantum entanglement 
 is an essential ingredient of quantum networks \cite{kimble2008quantum, hucul2015modular}.
Although optical photons were recently used to entangle solid-state qubits over long-distances\cite{hensen2015loophole,delteil2015generation},  
optically-mediated entanglement of solid-state qubits in a single nanophotonic device has not yet been achieved. 

\begin{figure}
\begin{center}
		\includegraphics[width=.65\linewidth]{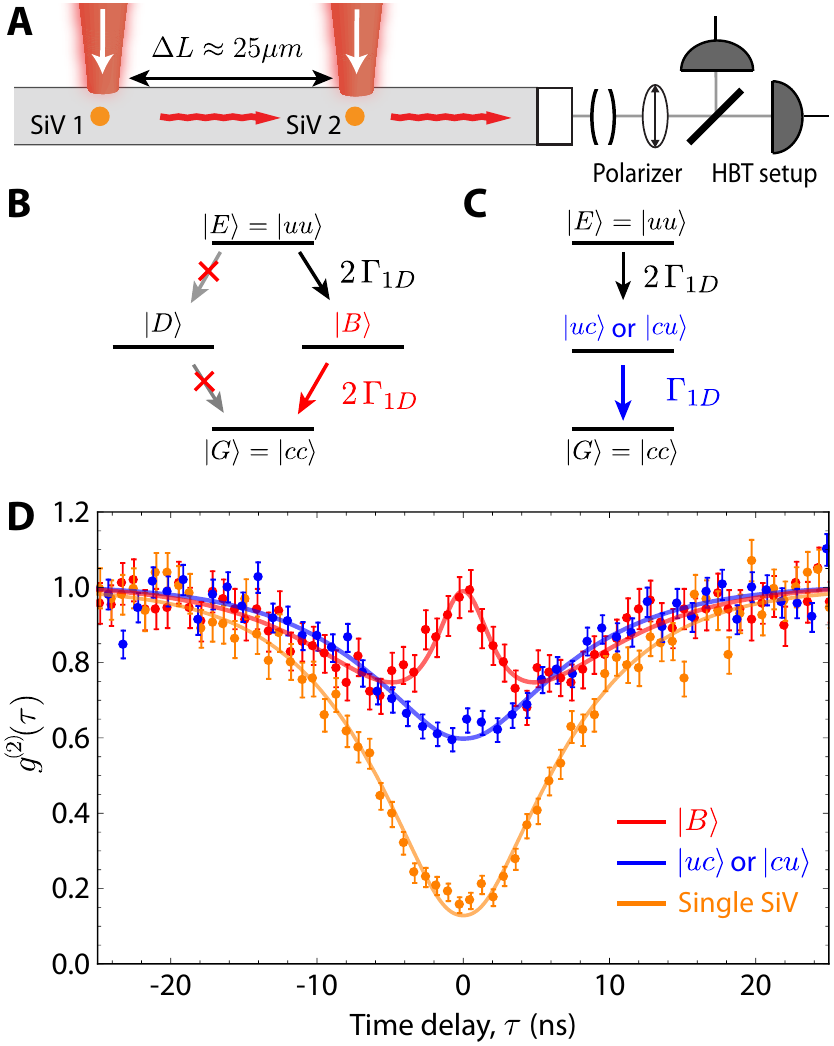}
\end{center}
		\caption{
	\textbf{Two-SiV entanglement in a diamond nanophotonic waveguide. }
\textbf{(A)} Two SiV centers in a diamond waveguide are excited from free space using separate laser fields. The scattered photons are sent through a polarizer to ensure polarization indistinguishability and detected in a Hanbury Brown-Twiss (HBT) setup. 
\textbf{(B)}  After detection of an indistinguishable photon, the two SiVs are in the entangled state $|B\rangle$ which decays at the collectively enhanced rate of $2~\Gamma_{1D}$ into the waveguide.
\textbf{(C)} After detection of a distinguishable photon, the two SiVs are in a classical mixture of states $|uc\rangle$ and $|cu\rangle$ which decays at a rate $\Gamma_{1D}$ into the waveguide.  
\textbf{(D)} Intensity autocorrelation for the waveguide Raman photons. Exciting only a single SiV yields $g^{(2)}_{single}(0)=0.16\,(3)$ for SiV1 (orange data), and $g^{(2)}_{single}(0)=0.16\,(2)$ for SiV2.
Blue: Both SiVs excited; Raman photons are spectrally distinguishable. 
Red: Both SiVs excited; Raman photons are tuned to be indistinguishable.
The observed contrast between the blue and the red curves at $g^{(2)}(0)$ is due to the collectively enhanced decay of state $|B\rangle$. The solid curves are fits to a model described in \cite{SOM}.
	}
	\label{fig:entanglement}
\end{figure}

We use two spatially-separated SiV centers
inside a one-dimensional diamond waveguide (Fig.~5A) and continuously excite each SiV on the $|u\rangle\rightarrow|e\rangle$ transition with a  separate laser. We utilize an efficient probabilistic entanglement generation scheme based on detection of indistinguishable photons\cite{cabrillo1999,delteil2015generation}. 
We match the frequency and intensity of the Raman emission of the two emitters by controlling the frequency and intensity of each laser field, resulting in  indistinguishable Raman photons scattered by two SiV centers. 
The generation and verification of the entangled state using photon correlations can be understood in terms of the level diagrams involving the metastable states $|u\rangle$ and $|c\rangle$ as shown in Figs.~5B and 5C\cite{dicke1954,wiegner2015,mlynek2014observation}. 
For two independent emitters in state $|u\rangle$, each SiV scatters Raman photons to the waveguide at a rate $\Gamma_{1D}$. 
When the Raman transitions of the two SiVs are tuned into resonance with each other,
it is fundamentally impossible to distinguish which of the two emitters produced a waveguide photon. Thus, detection of an indistinguishable single photon leaves the two SiV centers prepared in the entangled state $|B\rangle=(|cu\rangle + e^{i\phi}|uc\rangle)/\sqrt{2}$ \cite{cabrillo1999,wiegner2015} 
(Fig.~5B), where $\phi$ is set by the propagation phase between emitters spaced by $\Delta L$ and the relative phase of the driving lasers which is constant in each experimental run (Fig.~S9).
This state is a two-atom superradiant Dicke state with respect to the waveguide mode, independent of the value of $\Delta L$\cite{wiegner2015}. 
This implies that even though there is only a single excitation stored in the state $|B\rangle$, it will scatter Raman photons at a rate $2\Gamma_{1D}$ that is twice the scattering rate of a single emitter.
This enhanced emission rate into the waveguide mode can be used as a signature of entanglement.

Entanglement generation is verified using photon correlation measurements (Fig.~5, \cite{wiegner2015}).  We excite the Raman emission of two SiV centers with continuous lasers and observe photon correlations measured in the waveguide mode. 
If the Raman transitions of the two SiVs are not tuned into resonance, the photons are distinguishable and the detection of the first photon prepares the system in a statistical mixture of states $|uc\rangle$ and $|cu\rangle$ (Fig.~5C). After the first photon detection, these states scatter photons at the single emitter rate of $\Gamma_{1D}$ resulting in the measured $g^{(2)}_{dist}(0)=0.63\,(3)$ (blue curve in Fig.~5D) close to the conventional limit associated with two single photon emitters $g^{(2)}_{dist}(0)=0.5$ \cite{sipahigil2014indistinguishable,SOM}. 
Alternatively, if the Raman transitions of the two SiVs are tuned instead into resonance with each other,  a novel superradiant feature  is observed in photon correlations around zero time delay with  $g^{(2)}_{ind}(0)=0.98\,(5)$ (red curve in Fig.~5D). The observed interference peak at short time delays results from the collective emission enhancement associated with the entangled state $|B\rangle$ which enhances the probability of the second photon emission by a factor of two, resulting, in the ideal limit,  in $g^{(2)}_{ind}(0)=1$ (similar to photon correlations of laser light, see\cite{SOM}). 

The visibility of the interference  signal in photon correlation measurements in Fig.~5D can be used to evaluate a lower bound on the conditional entanglement fidelity $F=\langle B|\rho|B\rangle$\cite{SOM}. 
For experimental runs where we detect a photon coincidence within the interference window (rate $\sim\SI{0.5}{\Hz}$), 
we obtain a lower bound on the conditional fidelity of $F\geq 82(7)\%$.
This conditional fidelity is primarily limited by laser leakage and scattering from nearby SiVs that yield false detection events\cite{SOM}. Since it is evaluated upon the detection of two photons, it is not affected by imperfect state initialization and photon losses.
Our  measurements also demonstrate entanglement generation after a single Raman photon detection (rate $\sim\SI{30}{\kHz}$). 
As discussed in \cite{SOM}, using photon correlation data and steady state population of SiV orbital states, we find that a single photon detection results in an entangled two SiV state with positive concurrence $\mathcal{C}>0.090\,(0.024)$, which is limited by imperfect initialization in state $|uu\rangle$.  
This single photon detection approach can be extended to efficiently create high-fidelity, heralded entanglement,
by initializing the two SiVs and operating in the pulsed regime\cite{cabrillo1999,delteil2015generation}. 

The width of the interference signal in Fig.~5D can be used to extract a lifetime  $T_2^*\approx\SI{2.5}{ns}$ of the entangled state $|B\rangle$. The main contribution limiting this lifetime is imperfect spectral tuning of the two Raman photons from the two SiVs, resulting in a relative frequency detuning $\delta$. In this regime, detection of a first photon results in a state $|\psi (\tau)\rangle=(|cu\rangle+e^{i(\phi -2\pi \delta \tau)}|uc\rangle)/\sqrt{2}$  that oscillates at frequency $\delta$ between states $|B\rangle$ and the subradiant state $|D\rangle=(|cu\rangle - e^{i\phi}|uc\rangle)/\sqrt{2}$. Since $|D\rangle$ does not couple to the waveguide mode due to destructive interference, fluctuations in $\delta$ over different realizations result in decay of  the collectively enhanced signal (central peak in red curve in Fig.~5D).
In our experiment, this relative linewidth is limited by the procedure used to stabilize the Raman emission frequencies and could be improved by using a stable narrowband reference cavity.  Note that under the continuous driving used in our experiment, the optical pumping rate out of state $|u\rangle$  
determines the upper limit on the coherence time of state $|B\rangle$, an effect that can also be circumvented using a pulsed excitation scheme. 

\noindent \textbf{Outlook. }
In our experiments, control over the SiV orbital states is limited by finite phonon occupation at \SI{4}{\kelvin}, which causes relaxation between the metastable orbital states $|u\rangle$ and $|c\rangle$ and limits their coherence times to less than \SI{50}{ns}. 
Millisecond-long coherence times should be achievable by operating at temperatures below 300 mK or engineering the phononic density of states to suppress phonon relaxation processes\cite{jahnke2015electron,rogers2014all}. Even longer-lived quantum memories 
can potentially be obtained by storing the qubit in the ${}^{29}$Si nuclear spin ancilla \cite{rogers2014all}.

The demonstrated cooperativity in our nanocavity experiment is lower than the theoretical estimate based on an ideal two-level emitter optimally positioned in a cavity\cite{SOM}. 
The discrepancy is likely due to a combination of factors including imperfect spatial and polarization alignment, phonon broadening\cite{jahnke2015electron}, finite quantum efficiency\cite{riedrich2014deterministic}, the branching ratio of the transition and residual spectral diffusion\cite{evans2016narrow}. These imperfections also limit the collection efficiencies obtained in our waveguide experiments. 
Despite uncertainties in individual contributions, operation at lower temperatures and improved nanophotonic designs using slow-light waveguides\cite{Javadi2015} and cavities with higher $Q/V$ ratios 
should enable spin-photon interfaces with very high cooperativity $C\gg1$. 
Furthermore, the efficient fiber-diamond waveguide coupling can be improved to exceed $95\%$ efficiency \cite{tiecke2015efficient}.
The combination of these improvements should enable near-deterministic spin-photon interfaces operating at \SI{}{GHz} bandwidth.

Our work demonstrates key ingredients required for realizing on-chip quantum networks, including 
realization of single-photon switches, generation of tunable single photons  and entanglement of solid-state emitters.
These demonstrations open up new possibilities for realizing large-scale systems involving multiple emitters strongly interacting via photons. 
Our fabrication approach can be used to create systems involving many coupled emitters per cavity as well as arrays of multiple atom-cavity nodes. 
Using Raman tuning techniques, the cavity mode can be used as a quantum bus to generate complex multi-photon states with controlled interactions\cite{gonzalez2015deterministic}, as well as deterministic entanglement and tunable interactions between individually addressable emitters\cite{majer2007coupling,imamoglu1999quantum}. 
Because of the large cavity bandwidth, more than ten coupled emitters can be addressed independently in a single cavity to form complex quantum nodes that efficiently couple to a fiber network. 
Such a system can be used to implement robust quantum gates for either photonic or spin qubits with integrated error detection and correction \cite{borregaard2015heralded},
paving the way for the realization of integrated quantum networks with applications such as long-distance quantum communication\cite{kimble2008quantum, chang2014quantum,ladd2010quantum}. 

\baselineskip8pt 
\bibliography{SiVbib}

\begin{thebibliography}{10}

\bibitem{kimble2008quantum}
H.~J. Kimble, {\it Nature\/} {\bf 453}, 1023 (2008).

\bibitem{chang2014quantum}
D.~E. Chang, V.~Vuleti{\'c}, M.~D. Lukin, {\it Nat. Photon.\/} {\bf 8}, 685
  (2014).

\bibitem{ladd2010quantum}
T.~D. Ladd, {\it et~al.\/}, {\it Nature\/} {\bf 464}, 45 (2010).

\bibitem{tiecke2014nanophotonic}
T.~Tiecke, {\it et~al.\/}, {\it Nature\/} {\bf 508}, 241 (2014).

\bibitem{reiserer2014quantum}
A.~Reiserer, N.~Kalb, G.~Rempe, S.~Ritter, {\it Nature\/} {\bf 508}, 237
  (2014).

\bibitem{englund2007controlling}
D.~Englund, {\it et~al.\/}, {\it Nature\/} {\bf 450}, 857 (2007).

\bibitem{Javadi2015}
A.~Javadi, {\it et~al.\/}, {\it Nat. Commun.\/} {\bf 6}, 8655 (2015).

\bibitem{sun2016quantum}
S.~Sun, H.~Kim, G.~S. Solomon, E.~Waks, {\it Nat. Nanotechnol.\/}  (2016).

\bibitem{mlynek2014observation}
J.~Mlynek, A.~Abdumalikov, C.~Eichler, A.~Wallraff, {\it Nat. Commun.\/} {\bf
  5} (2014).

\bibitem{lodahl2015interfacing}
P.~Lodahl, S.~Mahmoodian, S.~Stobbe, {\it Rev. Mod. Phys.\/} {\bf 87}, 347
  (2015).

\bibitem{badolato2005deterministic}
A.~Badolato, {\it et~al.\/}, {\it Science\/} {\bf 308}, 1158 (2005).

\bibitem{faraon2012coupling}
A.~Faraon, C.~Santori, Z.~Huang, V.~M. Acosta, R.~G. Beausoleil, {\it Phys.
  Rev. Lett.\/} {\bf 109}, 033604 (2012).

\bibitem{riedrich2014deterministic}
J.~Riedrich-M{\"o}ller, {\it et~al.\/}, {\it Nano Lett.\/} {\bf 14}, 5281
  (2014).

\bibitem{hepp2014electronic}
C.~Hepp, {\it et~al.\/}, {\it Phys. Rev. Lett.\/} {\bf 112}, 036405 (2014).

\bibitem{rogers2014multiple}
L.~J. Rogers, {\it et~al.\/}, {\it Nat. Commun.\/} {\bf 5}, 4739 (2014).

\bibitem{sipahigil2014indistinguishable}
A.~Sipahigil, {\it et~al.\/}, {\it Phys. Rev. Lett.\/} {\bf 113}, 113602
  (2014).

\bibitem{evans2016narrow}
R.~E. Evans, A.~Sipahigil, D.~D. Sukachev, A.~S. Zibrov, M.~D. Lukin, {\it
  Phys. Rev. Appl.\/} {\bf 5}, 044010 (2016).

\bibitem{burek2014high}
M.~J. Burek, {\it et~al.\/}, {\it Nat. Commun.\/} {\bf 5}, 5718 (2014).

\bibitem{burek2012free}
M.~J. Burek, {\it et~al.\/}, {\it Nano Lett.\/} {\bf 12}, 6084 (2012).

\bibitem{SOM}
See Supplementary Material, available as an ancillary file on the arXiv, for
  materials and methods.

\bibitem{jahnke2015electron}
K.~D. Jahnke, {\it et~al.\/}, {\it New J. Phys.\/} {\bf 17}, 043011 (2015).

\bibitem{chang2007single}
D.~E. Chang, A.~S. S{\o}rensen, E.~A. Demler, M.~D. Lukin, {\it Nat. Phys.\/}
  {\bf 3}, 807 (2007).

\bibitem{duan2004scalable}
L.-M. Duan, H.~J. Kimble, {\it Phys. Rev. Lett.\/} {\bf 92}, 127902 (2004).

\bibitem{imamoglu1999quantum}
A.~Imamo\u{g}lu, {\it et~al.\/}, {\it Phys. Rev. Lett.\/} {\bf 83}, 4204
  (1999).

\bibitem{rice1988single}
P.~R. Rice, H.~J. Carmichael, {\it IEEE J. Quantum Elect.\/} {\bf 24}, 1351
  (1988).

\bibitem{pingault2014all}
B.~Pingault, {\it et~al.\/}, {\it Phys. Rev. Lett.\/} {\bf 113}, 263601 (2014).

\bibitem{rogers2014all}
L.~J. Rogers, {\it et~al.\/}, {\it Phys. Rev. Lett.\/} {\bf 113}, 263602
  (2014).

\bibitem{becker2016ultrafast}
J.~N. Becker, J.~G{\"o}rlitz, C.~Arend, M.~Markham, C.~Becher, {\it arXiv
  preprint arXiv:1603.00789\/}  (2016).

\bibitem{tiecke2015efficient}
T.~Tiecke, {\it et~al.\/}, {\it Optica\/} {\bf 2}, 70 (2015).

\bibitem{kuhn2002deterministic}
A.~Kuhn, M.~Hennrich, G.~Rempe, {\it Phys. Rev. Lett.\/} {\bf 89}, 067901
  (2002).

\bibitem{fernandez2009optically}
G.~Fernandez, T.~Volz, R.~Desbuquois, A.~Badolato, A.~Imamoglu, {\it Phys. Rev.
  Lett.\/} {\bf 103}, 087406 (2009).

\bibitem{he2013indistinguishable}
Y.~He, {\it et~al.\/}, {\it Phys. Rev. Lett.\/} {\bf 111}, 237403 (2013).

\bibitem{sweeney2014cavity}
T.~M. Sweeney, {\it et~al.\/}, {\it Nat. Photon.\/} {\bf 8}, 442 (2014).

\bibitem{hucul2015modular}
D.~Hucul, {\it et~al.\/}, {\it Nat. Phys.\/} {\bf 11}, 37 (2015).

\bibitem{hensen2015loophole}
B.~Hensen, {\it et~al.\/}, {\it Nature\/} {\bf 526}, 682 (2015).

\bibitem{delteil2015generation}
A.~Delteil, {\it et~al.\/}, {\it Nat. Phys.\/}  (2015).

\bibitem{cabrillo1999}
C.~Cabrillo, J.~Cirac, P.~Garcia-Fernandez, P.~Zoller, {\it Phys. Rev. A\/}
  {\bf 59}, 1025 (1999).

\bibitem{dicke1954}
R.~H. Dicke, {\it Phys. Rev.\/} {\bf 93}, 99 (1954).

\bibitem{wiegner2015}
R.~Wiegner, S.~Oppel, D.~Bhatti, J.~von Zanthier, G.~Agarwal, {\it Phys. Rev.
  A\/} {\bf 92}, 033832 (2015).

\bibitem{gonzalez2015deterministic}
A.~Gonz{\'a}lez-Tudela, V.~Paulisch, D.~E. Chang, H.~J. Kimble, J.~I. Cirac,
  {\it Phys. Rev. Lett.\/} {\bf 115}, 163603 (2015).

\bibitem{majer2007coupling}
J.~Majer, {\it et~al.\/}, {\it Nature\/} {\bf 449}, 443 (2007).

\bibitem{borregaard2015heralded}
J.~Borregaard, P.~K{\'o}m{\'a}r, E.~Kessler, A.~S. S{\o}rensen, M.~Lukin, {\it
  Phys. Rev. Lett.\/} {\bf 114}, 110502 (2015).

\end{thebibliography}
\baselineskip12pt 

\bibliographystyle{Science}
\noindent\textbf{Acknowledgements.} We thank D.\,J.\ Twitchen and M.\ Markham from Element Six Inc.\ for providing the diamond substrates and K.~De Greve and M.\,L.~Goldman for experimental help. 
Financial support was provided by the NSF, the Center for Ultracold Atoms,
the
AFOSR MURI, the DARPA QuINESS program, the ARL, the NSF GRFP (R.\,E.), the Carlsberg Foundation (J.\,B.) and the Harvard Quantum Optics Center (M.\,J.\,B.\ and H.\,A.).
Devices were fabricated at the Harvard Center for Nanoscale Systems supported under NSF award ECS-0335765. 
Ion implantation was performed at Sandia National Laboratories (SNL) with support from the Center for Integrated Nanotechnologies, an Office of Science (SC) facility operated for the U.S.~DOE SC (contract DE-AC04-94AL85000) by the Sandia Corporation, a subsidiary of Lockheed Martin.
\baselineskip16pt 

\end{document}